\documentstyle[twocolumn,epsbox]{jpsj}

\def\runtitle{
A Possible Mechanism of the Pseudogap in Organic Superconductor
Based on the Superconducting Fluctuation}

\def\runauthor{Takanobu {\sc Jujo} and Kosaku {\sc Yamada}}

\title
{
A Possible Mechanism of the Pseudogap in Organic Superconductor
Based on the Superconducting Fluctuation}

\author
{
Takanobu {\sc Jujo}\footnote{E-mail: jujo@ton.scphys.kyoto-u.ac.jp}
and Kosaku {\sc Yamada}
}

\inst
{
Department of Physics, Kyoto University, Kyoto 606-8502
}

\recdate
{March 19, 1999
}

\abst
{
The explanation of the anomalous behavior 
in $\kappa$-type (BEDT-TTF)$_2$X
which was revealed by the nuclear magnetic resonance 
experiments 
is presented. 
We calculate the electronic properties 
by using the one-loop approximation 
for the superconducting fluctuation. 
The decrease of the density of states due to the large damping 
effect is obtained and the one-particle 
spectrum has a pseudogap structure around the Fermi level. 
It is found that these behaviors are peculiar to the 
quasi two-dimensional system and are 
enhanced in the incoherent metal. 
The similarity between the cuprates and this compounds 
is discussed and an experiment to check this proposal 
is suggested. 
}

\kword
{organic superconductor, superconducting fluctuation, 
pseudogap, one-particle spectrum, 
quasi two-dimensional system, incoherent metal}

\begin{document}
\maketitle

Organic compounds are known to show 
various phases including superconductivity 
under controlled temperature and pressure.~\cite{rf:1}
In these compounds 
$\kappa$-type (BEDT-TTF)$_2$X
(abbreviated as $\kappa$-(ET)$_2$X hereafter, and \\
X=Cu(NCS)$_2$, 
Cu[N(CN)$_2$]Br, Cu[N(CN)$_2$]Cl) 
has attracted much attention because it has the highest 
superconducting transition temperature. 
The characteristic properties of $\kappa$-(ET)$_2$X are 
that the two-dimensional tight-binding approximation well 
explains the Shubnikov-de Haas experiments~\cite{rf:2}
and the superconducting phase 
neighbors the antiferromagnetic phase 
under a controlled pressure.~\cite{rf:3} 
About the properties of the superconductivity 
of $\kappa$-(ET)$_2$X, 
the nuclear magnetic resonance
(NMR)~\cite{rf:4} and the specific heat~\cite{rf:5} 
experiments of $\kappa$-(ET)$_2$X suggest that 
the superconducting gap has line nodes. 
Above the superconducting transition 
temperature
($T_{\rm c}$), 
$\kappa$-(ET)$_2$X shows the anomalous properties, 
which were revealed by the NMR experiments.~\cite{rf:6} 
The striking character of the anomalous properties 
is that the uniform magnetic susceptibility 
and $1/T_1T$ ($T_1$ is the spin lattice relaxation 
rate and $T$ is temperature) are not independent of 
temperature dependence but decrease 
with decreasing temperature below $T \simeq 50{\rm K}$. 
(This phenomenon may be called the spin gap or 
the pseudogap.) 

Theoretically, the importance of the electron correlation 
in $\kappa$-(ET)$_2$X is suggested because 
its dimerized structure makes the system half-filled.~\cite{rf:7} 
The mechanism of superconductivity 
of $\kappa$-(ET)$_2$X is investigated 
using several approaches like 
the fluctuation exchange approximation (FLEX),$^{8, 9)}$
the third-order perturbation theory (TOPT)~\cite{rf:10} 
and the quantum Monte Carlo simulation.~\cite{rf:11} 
From these studies, it is found that 
this superconductivity is a spin fluctuation 
mediated one and the symmetry of the Cooper pair 
is $d$-wave in this compound. 
However, for the anomalous properties above $T_{\rm c}$, 
there has been no theoretical proposal 
about the origin of these behaviors until now, 
and the aim of this paper is to propose 
a possible mechanism of the (pseudo-)spin gap 
in $\kappa$-(ET)$_2$X. 

$\kappa$-(ET)$_2$X has a strong anisotropy in the conductivity,~\cite{rf:12} 
that is, it can be considered that 
this compound is a quasi two-dimensional system. 
From FLEX and TOPT, 
it is also known that $T_{\rm c}$ 
of $\kappa$-(ET)$_2$X is rather high when 
$T_{\rm c}$ is scaled with the bandwidth. 
Regarding the properties of $\kappa$-(ET)$_2$X stated above, 
it can be considered that the superconducting fluctuation 
plays an essential role in the appearance of the (pseudo-)
spin gap. 

There are many theories on the pseudogap 
in cuprates on the basis of the superconducting fluctuation, 
but these calculations are restricted 
to a two-dimensional (2D) case$^{13, 14)}$
and the effect of the dimensionality has not been considered yet. 

In this Letter, we investigate the electronic properties 
for several values of the anisotropy parameter 
and for those of the attractive force by using the one-loop 
approximation for the superconducting fluctuation. 
From the nature of the one-loop calculation, 
our interest doesn't lie in the effect of the fluctuation on 
$T_{\rm c}$, but in the electronic properties above $T_{\rm c}$. 
(Therefore $T_{\rm c}$ is the mean field transition temperature 
in our discussion.) 
It is revealed that, owing to the superconducting, 
fluctuation the damping rate of quasiparticles increases 
and then the density of states (DOS) at the Fermi level 
decreases; this possibly explains the anomalous behavior. 
We also obtained the result that the one-particle spectrum 
has a pseudogap structure. This behavior is expected to be 
observed in the angle-resolved photoemission 
spectroscopy (ARPES) experiments. 
It is found that this pseudogap behavior 
is a characteristic property of a quasi 2D 
system such as $\kappa$-(ET)$_2$X. Furthermore, 
this behavior is enhanced in 
the incoherent metallic region where the 
spectrum is thermally broadened. 

To observe the effect of the dimensionality, we consider the following 
Hamiltonian, 
\begin{equation}
 {\cal H}=\sum_{{\mib k},\sigma}
\epsilon_{\mib k}c_{{\mib k},\sigma}^{\dag}c_{{\mib k},\sigma}
-\sum_{\mib k,{\mib k'},{\mib q}}g_{\mib k,{\mib k'}}
c_{{\mib k},\uparrow}^{\dag}c_{{\mib q-{\mib k}},\downarrow}^{\dag}
c_{{\mib q-{\mib k'}},\downarrow}c_{{\mib k'},\uparrow}.
\end{equation}
Here $\epsilon_{\mib k}$ is the noninteracting energy dispersion, 
\begin{equation}
  \epsilon_{ \mbox{\boldmath $k$}}=
-2t(\mbox{cos}k_x+\mbox{cos}k_y)-2t'\mbox{cos}(k_x+k_y)-2t_z{\rm cos}(2k_z)-\mu.
\end{equation} 
$\mu$ is the chemical potential, and we adopt the dimer model. 
Therefore $\mu$ is determined so as 
to fix the particle number per site to 1. 
We take $t$ as the unit of energy, and fix the ratio 
$t'/t$ to 0.70 as a realistic value.  
It should be noted that we include the inter-plane hopping term 
in the energy dispersion. The essential point to study 
the effect of the dimensionality, is not the form of this term 
but the ratio between $t,t'$ (intra-plane hopping) and 
$t_z$ (inter-plane hopping). 
We take the separable $d$-wave interaction 
as the attractive force, that is, 
$g_{{\mib k},{\mib k'}}=g\phi_{\mib k}\phi_{\mib k'}$, 
($\phi_{\mib k}={\rm cos}(k_x)-{\rm cos}(k_y)$). 

On the basis of this model, we take the superconducting 
fluctuation into account by the one-loop approximation, 
which is taking T-matrix up to the first order 
in calculating the self-energy.
The self-energy is written as the following form; 
\begin{equation} 
\Sigma(k)=
{\frac{T}{N}}\sum_{q}
T_{\mbox{\boldmath $k,k$}}(q)G_0(q-k). 
\end{equation}
Here, $N$ is the number of sites, 
$G_0$ is the bare Green's function and 
$T_{{\mib k},{\mib k'}}(q)$ 
is the T-matrix; 
\begin{equation}
T_{{\mib k},{\mib k'}}(q)=-\phi_{{\mib k}}
D(q)\phi_{{\mib k'}}, 
\end{equation}
where $D(q)$ is the boson propagator; 
\begin{equation}
D(q)=
\frac{1}{{\frac{1}{g}}-{\frac{T}{N}}\sum_{k}
\phi_{{\mib k}}^2
G_0(k)G_0(q-k)}.
\end{equation}
Here, $k$ is $(\mib k,\epsilon_n)$ ($\epsilon_n$ 
is the fermionic Matsubara frequency 
$\epsilon_n=(2n+1)\pi T$, n is integer), 
and $q$ is $(\mib q,\omega_m)$ ($\omega_m$ 
is the bosonic Matsubara frequency 
$\omega_m=2m\pi T$, m is integer). 
The numerical calculations are performed 
with $32 \times 32 \times 32$ meshes in the first 
Brillouin zone, and with 512 Matsubara frequencies. 
This number of frequencies ensures 
reliable results down to $T=0.003$. 

\begin{figure}
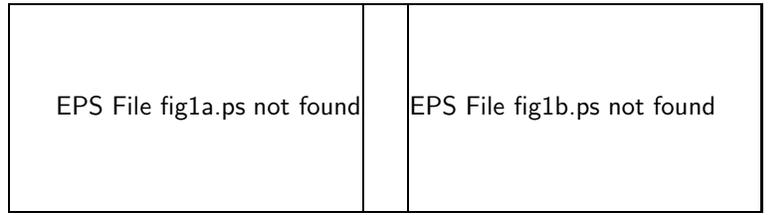

  \begin{minipage}[t]{.22\textwidth}
\epsfile{file=fig1a,height=3.5cm}
  \end{minipage}
  \hfill
  \begin{minipage}[t]{.22\textwidth}
\epsfile{file=fig1b,height=3.5cm}
  \end{minipage}
\caption{The self-energy 
for  $t_z=0.10$ ($T_{\rm c}=0.337$) 
and $t_z=1.00$ ($T_{\rm c}=0.195$). 
The values of $g$ and temperatures are shown in 
the figure. 
(a) The real part of the self-energy 
(b) The imaginary part of the self-energy. 
}
\label{fig:1}
\end{figure}

Now, we show the results obtained by our calculation. 
First, we see the effect of the dimensionality 
by calculating the electronic properties 
for different values of $t_z$. 
The calculated self-energies 
for the inter-plane transfer $t_z=0.10$ and $1.00$ 
are shown in Fig.~\ref{fig:1}. 
The momentum of the self-energies is fixed to 
the Fermi momentum situated 
on the line connecting $\Gamma$ point $(0,0,0)$ and 
$M$ point $(\pi,0,0)$. 
At this point called $A$ (see Fig. 5(a)), 
the attractive force takes the strongest value due
to the $d_{x^2-y^2}$-wave symmetry. 
For the isotropic case ($t_z=1.00$), the behavior of the self-energy 
is the same as that of the conventional Fermi liquid. 
That is, 
$-{\rm Im}\Sigma({\mib k},\omega)$ has a 
local minimum at $\omega =0 $ and the slope of
${\rm Re}\Sigma({\mib k},\omega)$ 
around $\omega=0$ is negative. 
These properties of the self-energy indicate the well-defined 
quasiparticle.~\cite{rf:15}
However, for the quasi 2D case ($t_z=0.10$), the behavior of the self-energy 
is opposite to that of the isotropic case. 
The peak of $-{\rm Im}\Sigma({\mib k},\omega)$ at $\omega=0$
indicates that the quasiparticle is not well-defined, 
and the positive slope of ${\rm Re}\Sigma({\mib k},\omega)$ 
around $\omega=0$ indicates that the spectral weight of quasiparticles 
near the Fermi surface decreases. Therefore the DOS at the 
Fermi level is expected to be suppressed. 
The fact that 
there exist these anomalous behaviors 
of the self-energy behind the precursor of the superconductivity 
was first pointed out by Jank\'o, {\it et al.}~\cite{rf:14} 
In addition to this fact, it is pointed out here that these 
anomalous properties are seen only in the quasi 2D 
system. 

\begin{figure}
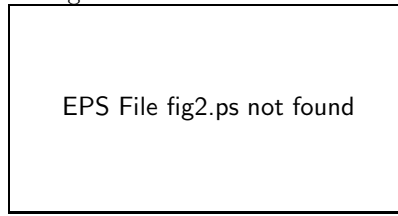

\epsfile{file=fig2,height=5.5cm}
\caption{The temperature dependences of the DOS 
at the Fermi level $\rho (0)$, and the 
absolute value of the  
imaginary part of the self-energy at the 
FS and the Fermi level 
$-{\rm Im}\Sigma({\mbox {\boldmath $k$}}_{\rm F},0)$. 
Here ${\mbox {\boldmath $k$}}_{\rm F}$ is at the point $A$. 
``ise'' in the figure means the imaginary part of the 
self-energy.
$T_{\rm c}$ is the same as in Fig. 1.}
\label{fig:2}
\end{figure}

To observe the relationship between the self-energy 
and the DOS, 
the temperature dependences of $\rho (0)$ (the DOS 
at the Fermi level) and 
$-{\rm Im}\Sigma({\mib k}_{\rm F},0)$ 
are shown in Fig.~\ref{fig:2}. 
It can be seen that for 
the quasi 2D case ($t_z=0.10$) $\rho (0)$ begins to decrease 
far above $T_{\rm c}$ with the increase of 
the damping rate of quasiparticles near the Fermi level, 
$-{\rm Im}\Sigma({\mib k}_{\rm F},0)$. 
On the other hand, for the isotropic case ($t_z=1.00$) 
$\rho (0)$ increases down to $T_{\rm c}$ 
because the system acquires coherency as the temperature 
is decreased, which is supported by the decrease of the 
damping rate of quasiparticles, $-{\rm Im}\Sigma({\mib k}_{\rm F},0)$. 
Although, as discussed below, two other factors are required 
for the quantitative comparison with experimental results, 
the decrease of $\rho (0)$ over a 
wide temperature range 
suggests that the precursor of the superconductivity 
can be a candidate mechanism which gives rise to the 
anomalous behavior in $\kappa$-(ET)$_2$X, because 
the quantities showing the anomalous behavior in the experiments 
are related to the DOS at the Fermi level. 

The above discussion is described analytically 
by assuming that the boson propagator 
is written in the following form; 

\begin{equation}
D({\mib q},\omega) \simeq 
\frac{1}{M-{\rm i}a_i\omega+\xi^2{\mib q}^2}.
\end{equation}
Here, $M$ is the mass term which 
describes how far the system is from 
the superconducting state and is written as 
$M=\frac{1}{g}-\sum_{\mib k}\phi_{\mib k}^2
\frac{{\rm tanh}\epsilon_{\mib k}/2T}{2\epsilon_{\mib k}}$, 
$M=0$ at $T=T_{\rm c}$. 

With the use of this form for the boson propagator, 
$-{\rm Im}\Sigma({\mib k}_{\rm F},0)$ 
can be written in the following form; 

\begin{equation} 
-{\rm Im}\Sigma({\mib k}_{\rm F},0) \simeq 
\frac{T\phi_{{\mib k}{\rm F}}^2}
{16 \pi v_{\rm F} \xi^2}{\rm log}(1+\frac{(\xi q_{\rm c})^2}{M}), 
\end{equation}
for a 3D system, and 

\begin{equation}
-{\rm Im}\Sigma({\mib k}_{\rm F},0) \simeq 
\frac{T\phi_{{\mib k}{\rm F}}^2}
{2 \pi v_{\rm F} \xi \sqrt{M}}{\rm tan}^{-1}(\frac{\xi q_{\rm c}}{\sqrt{M}}), 
\end{equation}
for a 2D system. 
Here, $q_{\rm c}$ is the cutoff momentum of the boson, and $v_{\rm F}$ is the velocity 
of the electron at the Fermi surface (FS). 

From these equations, it can be seen that for a 2D 
system $-{\rm Im}\Sigma({\mib k}_{\rm F},0)$ is large over 
a wide temperature region above $T_{\rm c}$. 
The interpolating formula between 3D and 2D cases 
can be written following the discussion by Takahashi,~\cite{rf:16} 

\begin{eqnarray}
-{\rm Im}\Sigma({\mib k}_{\rm F},0) \simeq
\frac{T\phi_{{\mib k}{\rm F}}^2}{16 \pi v_{\rm F} \xi^2 \epsilon}
{\rm log}(1+\frac{(\epsilon \xi q_c)^2}{M}) \nonumber\\
+\frac{T\phi_{{\mib k}{\rm F}}^2}{2 \pi v_{\rm F} \xi \sqrt{M}}
[{\rm tan}^{-1}(\frac{\xi q_c}{\sqrt{M}})
-{\rm tan}^{-1}(\frac{\epsilon \xi q_c}{\sqrt{M}})].
\end{eqnarray}
Here, $\epsilon \simeq t_z/t$ is the parameter which 
characterizes the degree of the anisotropy. 

We also calculated the one-particle spectrum 
($A({\mib k},\omega)=-{\rm Im}G({\mib k},\omega)/\pi$), 
and it is found that for the small inter-plane transfer $t_z$ 
the spectrum has a pseudogap structure around $\omega =0$. 
The relationship between $t_z$ and 
the temperature ($T_{\rm pg}$) 
at which the pseudogap opens at point $A$, 
is shown in Fig.~\ref{fig:3}. 
$T_{\rm pg}$ is obtained in a manner such that the one-particle
spectrum is calculated at intervals of 0.001 at a temperature 
which is less than one percent of $T_{\rm c}$, and $T_{\rm pg}$ is 
the temperature at which it is found to be suppressed 
around $\omega =0$. 
\begin{figure}
\epsfile{file=fig3,height=5.5cm}
\caption{The relationship between the anisotropy parameter 
($t_z$) and the temperature ($T_{\rm pg}$) 
at which the pseudogap opens in the 
one-particle spectrum at the point $A$. 
The value of $g$ is shown in the figure. 
$M$ is the mass-term defined above.}
\label{fig:3}
\end{figure}
It is seen that for the small 
inter-plane transfer $t_z$ the gap-like behavior can be seen 
at rather high temperatures above $T_{\rm c}$.~\cite{rf:17} 
Therefore, the pseudogap in the one-particle spectrum 
is expected to be found in $\kappa$-(ET)$_2$X which 
is a quasi 2D system. 
On the other hand, for large $t_z$, $T_{\rm c}$ is low 
because of the large bandwidth. Therefore, we studied 
the spectrum for large $g$ in the case of large $t_z$. 
However, we saw no gap-like behavior in this case either. 
Therefore we can conclude that the gap-like behavior 
is a characteristic property of the quasi 2D system. 
  
So far, we have fixed the magnitude of the 
attractive force $g$. 
Next, we will see whether the pseudogap behavior 
can be observed for all values of $g$. 
The calculations are performed by 
determining the value of $M$ for 
which the pseudogap opens for various values of $g$. 
Using this method, we can calculate the one-particle 
spectrum at the same point away from the superconducting 
state for various values of $g$. 
The results are shown in Fig.~\ref{fig:4}. 
\begin{figure}
\epsfile{file=fig4,height=5.5cm}
\caption{The relationship between the attractive force 
($g$) and the temperature ($T_{\rm pg}$) 
at which the pseudogap opens in the 
one-particle spectrum at point $A$.
The value of $t_z$ is shown in the figure.
$M$ is the mass-term.}
\label{fig:4}
\end{figure}
It can be seen that 
the pseudogap behavior is enhanced 
for large $g$. 
This result is interesting because the self-energy 
is independent of $g$ when the self-energy is 
written as a function of $M$ and therefore 
it is expected that the gap-like behavior 
appears for the same value of $M$. 
However, our results show that 
this assumption is incorrect. 
The behavior in Fig. 4 is explained by the 
following argument. 
The self-energy is written from eq. $(8)$, 
\begin{equation}
-{\rm Im}\Sigma({\mib k}_{\rm F},0) \propto
\sqrt{T_{\rm F}}(\frac{T}{T_{\rm F}})^2
\frac{1}{\sqrt{M}}.
\end{equation}
Here $T_{\rm F}$ is the Fermi degeneracy temperature. 
For large $g$, $T_{\rm c}$ is high and 
therefore the enhancement factor $(T/T_{\rm F})^2$
is large. 
Therefore, in the case of high $T_{\rm c}$, 
the damping effect is more enhanced even for rather 
large values of $M$ than in the case of low $T_{\rm c}$. 
In other words the thermally broadened spectrum 
is more sensitive to the superconducting fluctuation. 

This situation is considered to be realized 
because the superconductivity of $\kappa$-(ET)$_2$X 
appears only near the Mott-insulating phase and it is 
likely that the $T_{\rm F}$ is very low owing to 
the strong electron correlation. 
From the above discussion, we propose that point $A$ 
and its symmetry-related points of the Fermi surface 
must be destructed in this pseudogap region due to 
the strong fluctuation of the superconductivity. 
This will be checked by the ARPES experiments. 
The FS and the one-particle spectrum at point $A$ 
are shown in Fig.~\ref{fig:5}. 
\begin{figure}
  \begin{minipage}[t]{.22\textwidth}
\epsfile{file=fig5a,height=3.5cm}
  \end{minipage}
  \hfill
  \begin{minipage}[t]{.22\textwidth}
\epsfile{file=fig5b,height=3.5cm}
  \end{minipage}
\caption{(a) The Fermi surface for $t_z=0.00$. 
The meaning of point $A$ is the same as above. 
(b) The temperature dependence of the one-particle 
spectrum. The values of $g$ and $t_z$ are shown in 
the figure. In this case $T_{\rm c}$ is 
0.339.}

\label{fig:5}
\end{figure}
In Fig. 5(a), the FS of the single-band in our model 
is shown. The equivalence of this single band model 
and the two-band model (the hopping parameter $t$ 
takes two kinds of values) is discussed in ref. 8. 
From Fig. 5(b), it can be seen that the one-particle 
spectrum has a pseudogap structure above $T_{\rm c}$. 
This behavior disappears at the points of the FS 
away from point $A$ and its symmetry-related points because of 
the $d_{x^2-y^2}$-wave symmetry of the attractive force. 

Finally, we comment on the two factors which are not 
included in our discussion but which are required to be 
considered for an 
accurate quantitative description. 
One, is that we do not take into 
account the effect of the fluctuation in the determination 
of $T_{\rm c}$, i.e. the self-consistency requirement. 
For a quasi 2D system such as $\kappa$-(ET)$_2$X, 
the fluctuation plays an essential role in determining 
$T_{\rm c}$. However, our conclusion that the 
pseudogap is peculiar to the quasi 2D
system is not affected because our argument 
is basing on the mass term $M$ which measures 
the extent to which the system is away from the superconducting state. 
The other factor is the effect of the Coulomb repulsion. 
This is also important for obtaining accurate quantitative 
results in the system which neighbors the Mott insulating 
phase. However, if the adiabatic continuity is realized, 
the effect of the Coulomb repulsion 
is only the rescaling of various quantities, such as 
the Fermi degeneracy temperature. 
Justification of the adiabatic continuity in the metallic 
phase near the Mott-insulating phase is an important problem,~\cite{rf:18} 
and will be discussed elsewhere. 

In summary, we have calculated the electronic properties 
of the quasi 2D model 
by considering the superconducting fluctuation. 
The relationship between the DOS and the damping rate 
is shown and it is found that 
the decrease of the DOS is caused by the large damping rate 
which is enhanced by the superconducting fluctuation 
in the quasi 2D case. 
We have also calculated the one-particle spectrum, and 
the pseudogap behavior over a wide temperature 
range above $T_{\rm c}$ 
is obtained only for the quasi 2D case. 
It is also found that this behavior is 
characteristic of a rather incoherent metal 
which has a broad spectrum around the Fermi level 
and is sensitive to the fluctuation. 
These conditions for the appearance of the 
pseudogap can be applied to $\kappa$-(ET)$_2$X. 
Therefore, the anomalous properties 
found by the NMR experiments are considered to 
be caused by the superconducting fluctuation. 
The strong two-dimensionality and 
the incoherent metallic properties 
are also the properties of cuprates 
in the under-doped region, 
so our argument can be applied to this material 
and supports the expectation that 
the anomalous behaviors of the cuprates 
and $\kappa$-(ET)$_2$X have the same origin.$^{8, 19)}$
ARPES experiments are desired 
not only to check the validity of our proposal, 
but also to understand the property 
of pseudogap in $\kappa$-(ET)$_2$X more 
precisely. 

Numerical computation in this work was carried out at the 
Yukawa Institute Computer Facility.


\begin{thebibliography}{99}
\bibitem{rf:17} The bulge of $M$ at $t_z=0.3$ is explained by 
the following argument. For large $t_z$, point $A$ moves toward 
$(\pi,0)$, so the attractive force is large. Therefore 
the strong two-dimensionality at $t_z=0.3$ compared with 
$t_z \geq 0.5$ causes this result.
\end{thebibliography}
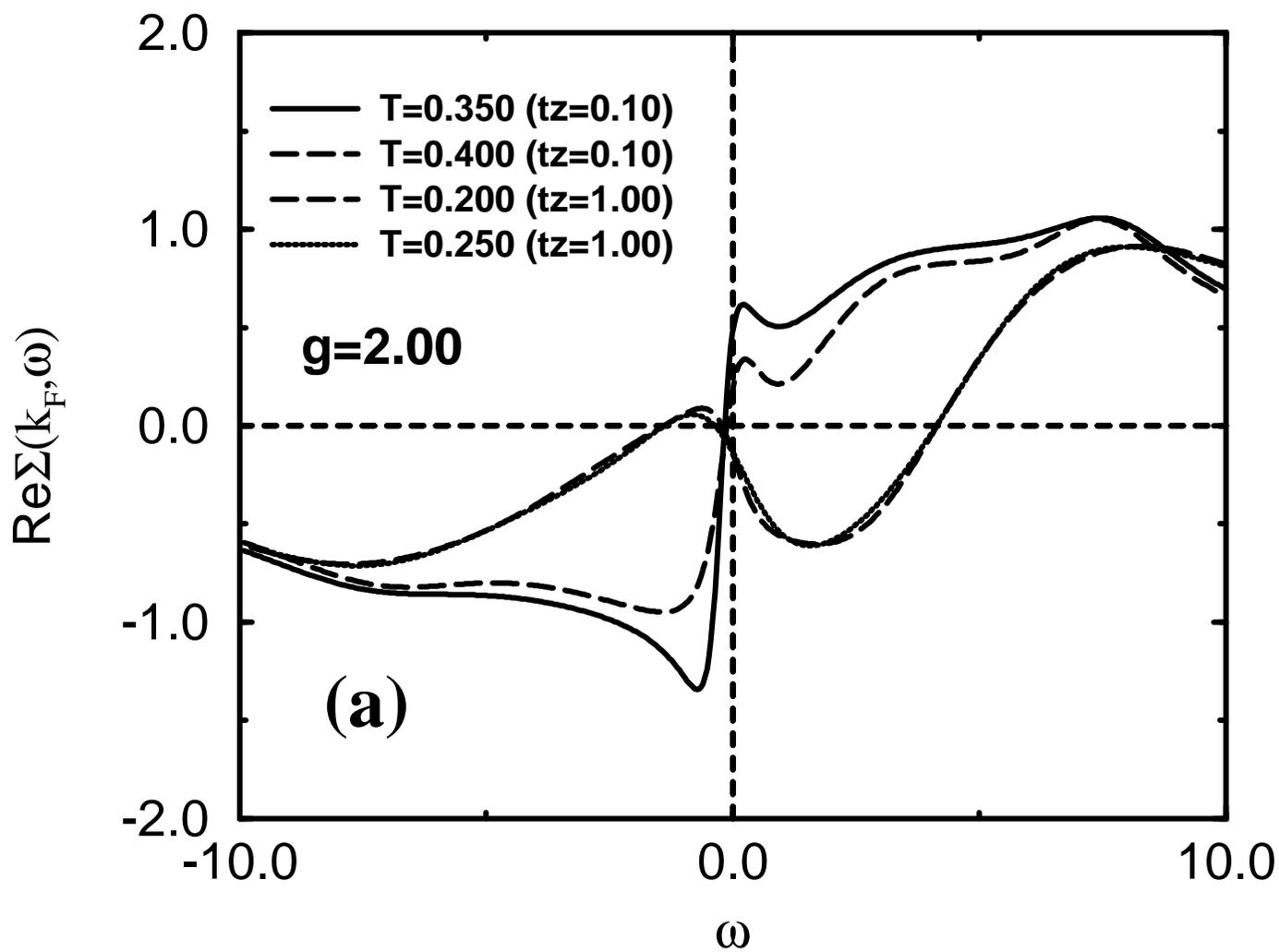
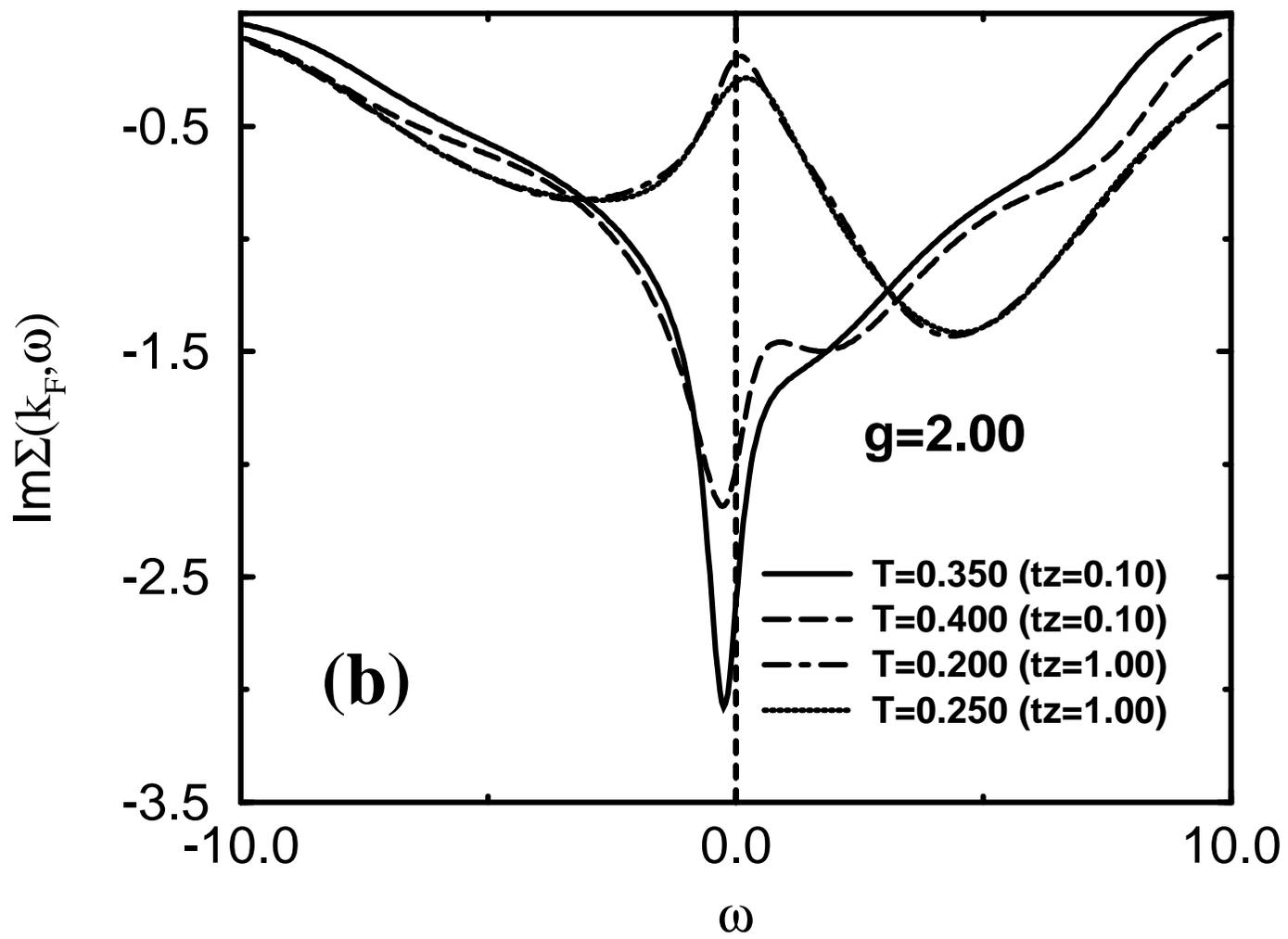
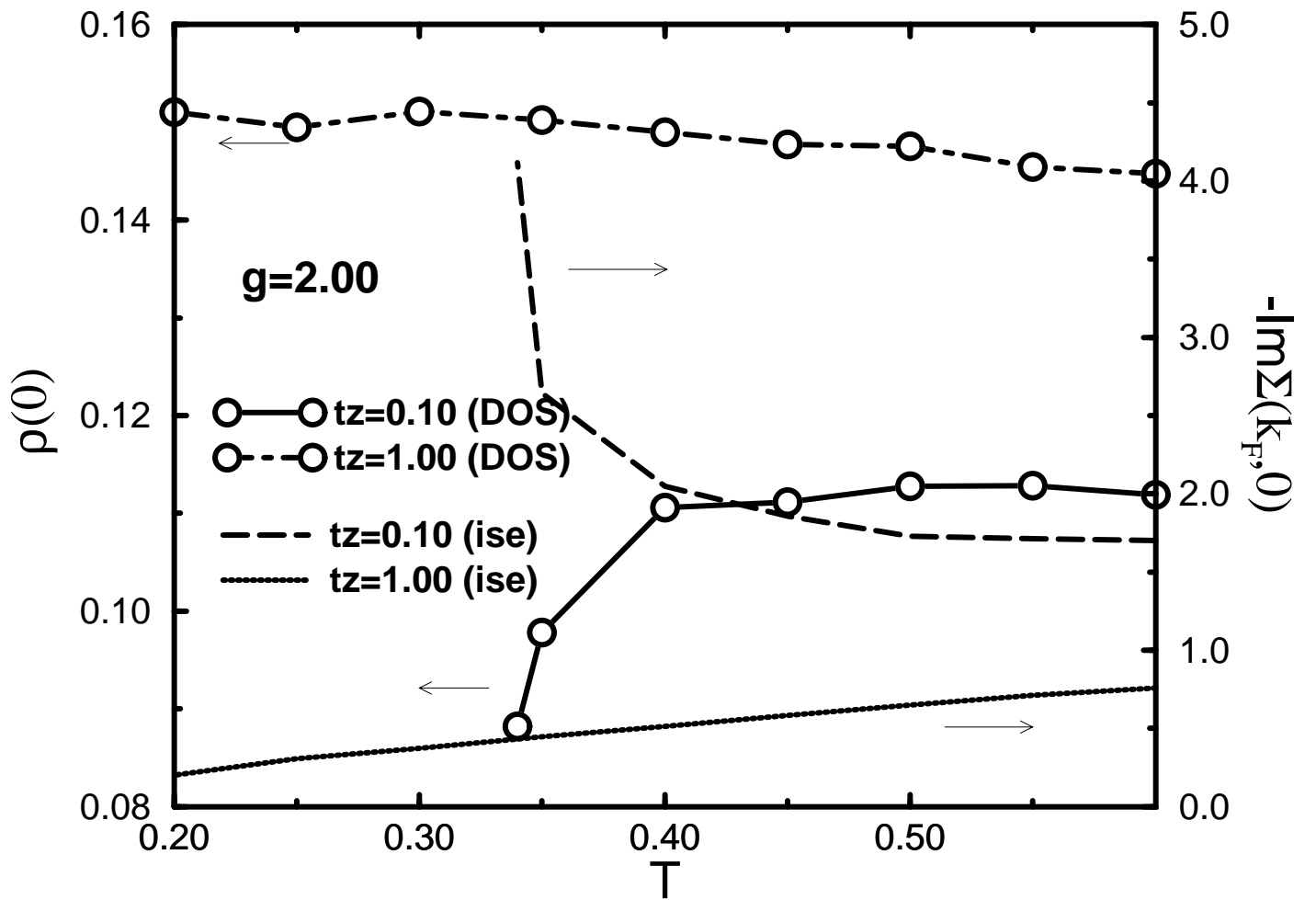
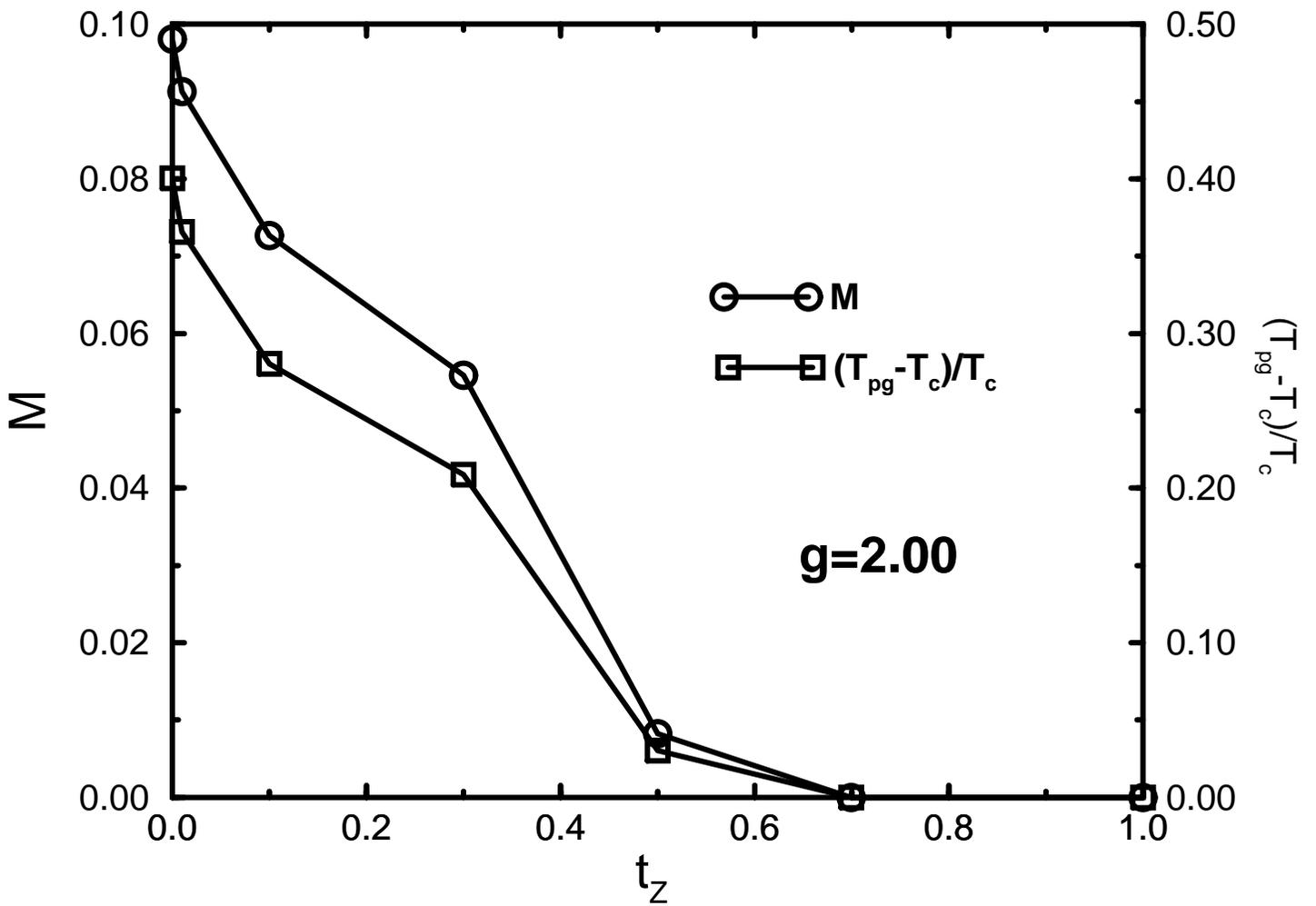
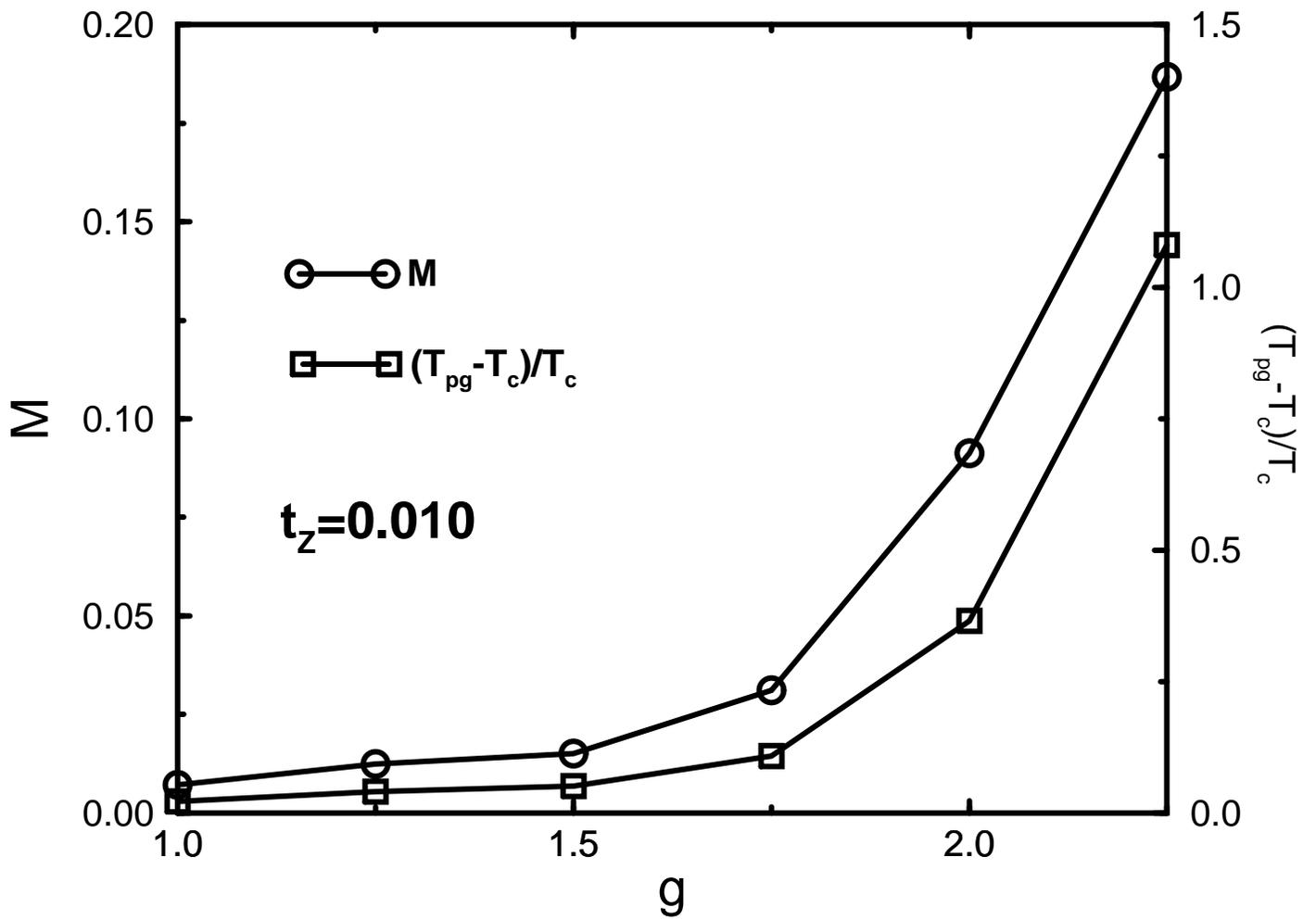
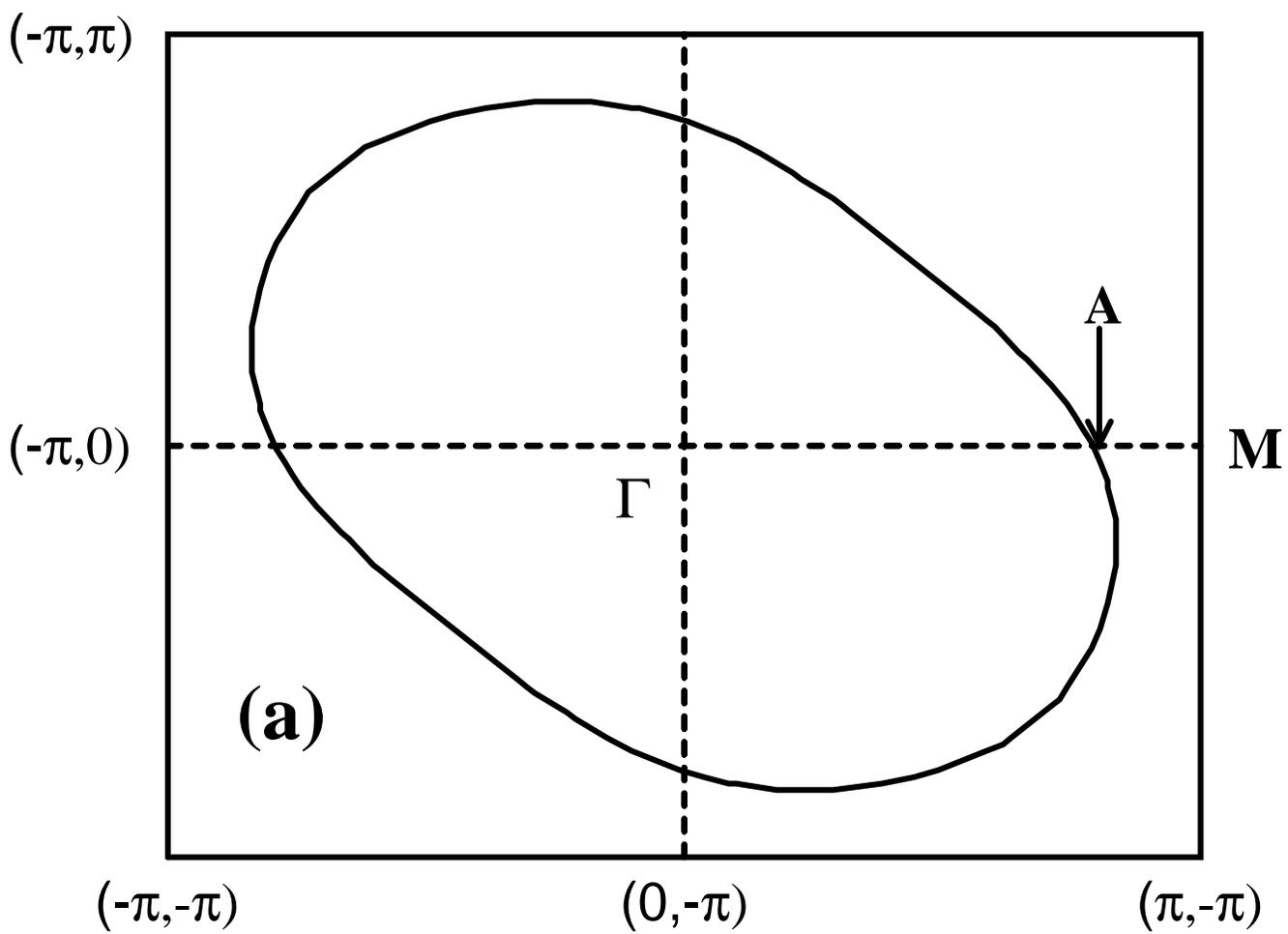
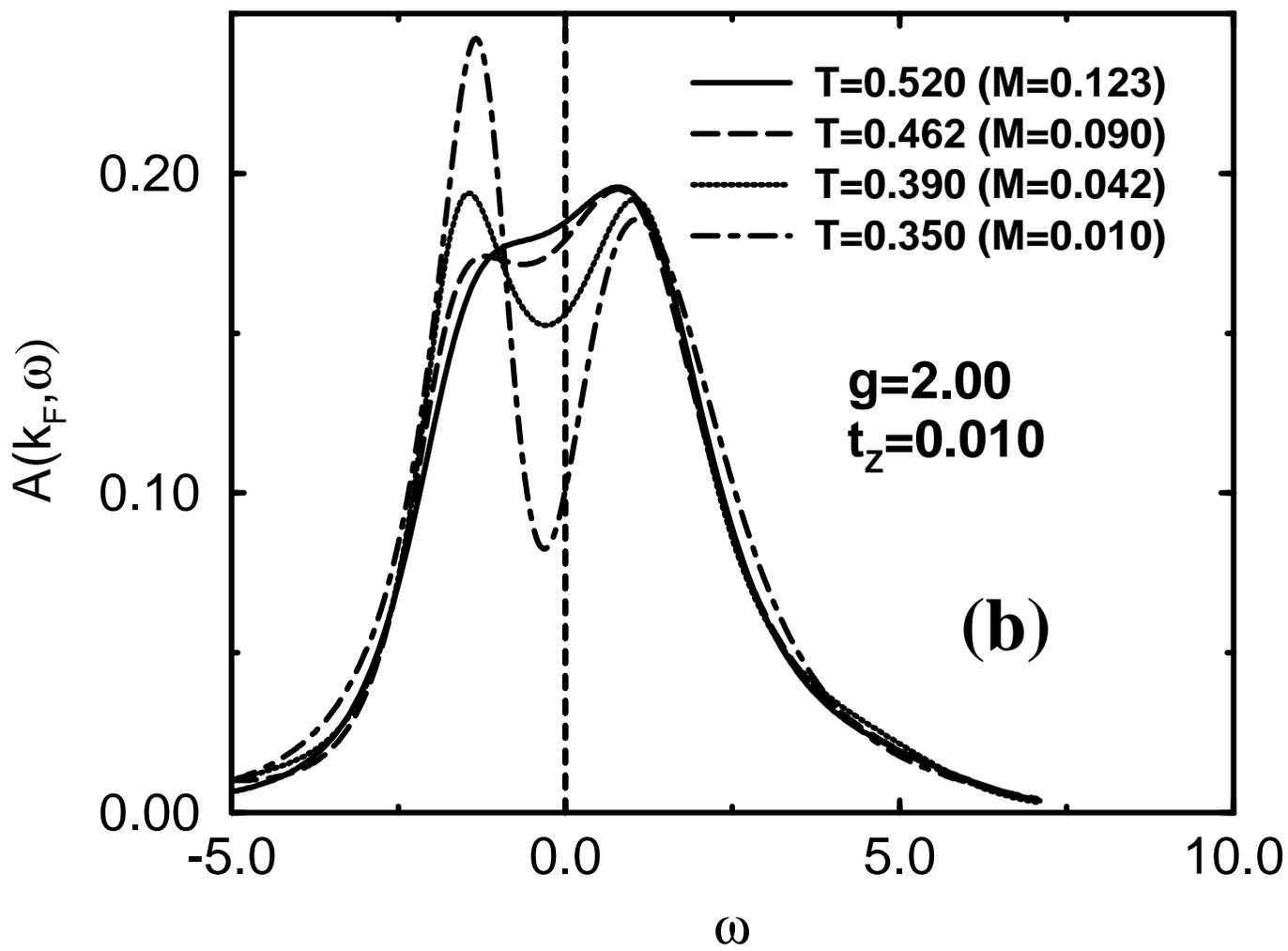
\end{document}